\documentstyle[12pt,epsf]{article}
\newcommand{\bi}{\bibitem}
\begin{document}
\baselineskip = 14pt
%\large
\begin{center}
{\LARGE\bf  The $p(d,p)d$ and $p(d,p)pn$ reactions as
a tool for the study of the short range
internal structure of the deuteron}\\
\vspace{0.5cm}
{\Large
V.P.~Ladygin$^*$}
\vspace{0.3cm}

{\em  LHE-JINR,
141980, Dubna, Moscow 
Region, Russia } \\ 

{$^*$ - E-mail : ladygin@sunhe.jinr.ru}\\
\end{center}

\vspace{0.3cm}

\begin{center}
{\large\bf Abstract}
\end{center}

\vspace{0.3cm}
\begin{quote}

In recent time the deuteron structure at short distances is
often treated from the point of view nonnucleonic degrees 
of freedom. In this paper
the measurements of
$T$-odd polarization observables using 
tensor polarized deuteron beam and polarized proton target or
proton polarimeter 
are proposed
to search the  quark configurations
inside the deuteron. 

\end{quote}

%\newpage

\section{Introduction}

Recent experimental results concerning the structure of the deuteron have
led to the speculations that manifestation of the quark-gluon degrees
of freedom are present even at relatively large distances between nucleons.
Measurements of the cross section of the
inclusive deuteron breakup  $A(d,p)X$ reaction on carbon
with the proton emitted at a
zero degree  \cite{deucr0} have shown the relatively broad shoulder at
internal nucleon momenta $k\sim 0.35$ GeV/c
in the deuteron defined in the light-cone dynamics
\cite{dirac}-\cite{fs1}. This enhancement has been
observed later
at different initial energies and for different
$A$ values of the target \cite{deucr}-\cite{t20br1}.
This shoulder could not be reproduced by the calculations
within relativistic impulse approximation (IA) using the
standard deuteron wave functions  \cite{paris,rsc,bonn}, as well as by
the inclusion of the rescattering corrections \cite{bert}.
Theoretical work of Kobushkin and Vizireva led to
the possibility of existing of a $6q$ admixture in the deuteron
wave function \cite{kob1}.
This $6q$ amplitude, arising from the $S$ configurations of
six quarks, must be added to the  $S$ component of the
standard deuteron wave
function (DWF) with a relative phase, $\chi$.
The fit of the experimental data  \cite{deucr} gave the
probability of the $6q$ configuration about $\sim 4\%$ and
relative angle $\chi$ $\sim 82^0$ and $ \sim 61^0$ for
Paris \cite{paris} and Reid Soft Core (RSC) \cite{rsc} $NN$ potentials, 
respectively.  Admixture of a $6q$ state of about $3.4\%$  was imposed 
also in \cite{ign} to describe the tail of momentum spectrum of the 
$^{12}C(d,p)X$ reaction \cite{deucr}.

One of the important features of this hybrid wave function is that an
additional $6q$ admixture masks the node of the $NN$ $S$ wave,
what drastically reflects on the behaviour of polarization observables.
For instance, the data on tensor analyzing power $T_{20}$ and
cross section in
inclusive deuteron breakup at zero degree and at $2.1$ GeV of the
initial energy obtained at Saclay \cite{t20br1} were explained
by the hybrid wave function
with $\sim 4\%$ of $|6q>$ configuration probability with $55^0$ of 
relative phase between $|6q>$ and $S$ component from RSC potential 
\cite{rsc}.

Recent measurements of tensor analyzing power $T_{20}$ for deuteron
inclusive breakup at $0^0$ performed in Saclay \cite{t20br1} and in 
Dubna \cite{t20dub,anomalon,ladygin} at different energies and for 
different targets have shown the strong deviation from the IA 
predictions at $k\ge 0.2$ GeV/c.  
The behaviour of the polarization 
transfer coefficient from vector polarized deuteron to proton 
$\kappa_0$ \cite{kappa0,kappa1,kappa2} also disagrees with the 
calculations using conventional DWFs at $k\ge 0.2$ GeV/c.  On the 
other hand, both $T_{20}$ and $\kappa_0$ data demonstrate a weak 
dependence on $A$ value of the target, as well as an approximate energy 
independence, i.e.  features of IA.  Considering of the mechanisms 
additional to IA \cite{lykasov,vina} can not explain the experimental 
data.

Most intriguing feature of the experimental data is that
tensor analyzing power $T_{20}$ in deuteron inclusive breakup and
deuteron--proton backward elastic scattering  
show at high internal momenta of proton
the same negative value $\sim -0.3\div -0.4$ \cite{anomalon,ladygin,dppddub},
incompatible with the predictions using any reasonable nucleon--nucleon
potential.
Various attempts were undertook to explain the $T_{20}$ data taking into
account the nonnucleon degrees of freedom in the deuteron.
An asymptotic negative limit  of $T_{20}$ was
obtained in framework of the QCD motivated approach \cite{kob_QCD}
based on reduced nuclear amplitude method \cite{brodsky}.
The results of calculations \cite{lykasov} with the hybrid DWF
\cite{efremov}  allowed to
describe satisfactory the $T_{20}$ data up to
$k\sim 1$ GeV/c \cite{anomalon}.
Recently the data on $T_{20}$ and $\kappa_0$ in the
$^{12}C(d,p)X$ reaction at $0^0$ were reasonably reproduced within a 
model which incorporates multiple scattering and Pauli principle at the 
quark level \cite{t20kob}.  The additional account of the negative 
parity nucleon resonance exchanges improves the accordance of 
calculations with the experimental data on $T_{20}$ in backward elastic 
$dp$ scattering \cite{yudin}.  The tensor analyzing power $A_{yy}$ in 
deuteron inclusive breakup obtained up to 600 MeV/c of a proton 
transverse momenta \cite{ayy} also disagrees with the calculations 
within  hard scattering model \cite{rhs} using conventional DWFs.
However, the sign of $A_{yy}$ at large 
proton momenta  as at a zero angle 
\cite{t20dub,anomalon,ladygin}
\footnote{At a zero angle emission angle $A_{yy} = -T_{20}/\sqrt{2}$}, 
as well as at a $\sim 90^0$ in the rest frame of the deuteron 
\cite{ayy} is the same as predicted by QCD motivated approach 
\cite{kob_QCD}.

These peculiarities of the experimental data and relative successful
attempts to describe them by the considering of the nonnucleon
degrees of freedom stimulate to measure additional polarization
observables crucial to the quark degrees of freedom
in the deuteron.

In our previous paper \cite{lad97}
we have considered the using of the polarized proton
target and proton polarimeter to study the deuteron structure
at short distances.
Here we propose to study $T$-odd polarization observables \cite{ohlsen}
in deuteron
exclusive breakup in the collinear geometry 
and $dp$ backward elastic scattering in order to identify the exotic 
$6q$ configurations inside the deuteron.

\section{Matrix elements of the $dp\to ppn$ and  $dp\to pd$ reactions}

In this section we analyze the polarization effects in
two processes : deuteron breakup in the strictly collinear geometry,
$d+p \to p(0^0) + p(180^0) +n$,  and
deuteron--proton backward elastic scattering, $d+p\to p+d$, using
the hybrid DWF with the complex $6q$ admixture.

This function can be presented in the momentum space
in the following form:
 
{\small
\begin{eqnarray}
\label{dwf}
\Phi_d({\bf{p}})= \frac{\it i}{\sqrt{2}} \frac 
{1}{\sqrt{4\pi}}\psi^{\alpha +}_{p} \left [\left ( U(p) 
({\vec{\sigma}}\cdot{\vec{\xi}})
 - \frac{W(p)}{\sqrt{2}} (3 (\hat{\bf 
p}{\vec{\xi}})({\vec{\sigma}}\hat{\bf p})-({\vec 
{\sigma}}\cdot{\vec{\xi}}))\right ) \sigma_y \right 
]_{\alpha\beta}\psi^{\beta +}_{n}, 
\end{eqnarray} 
}
where $\psi_p$ and 
$\psi_n$ are the proton and neutron spinors, respectively, $\vec{\xi}$ 
is the deuteron polarization vector, defined in a standard manner:  
\begin{eqnarray} 
{\vec{\xi}}_1 = -\frac{1}{\sqrt{2}} (1, {\it i}, 
0)~~~~~~~ {\vec{\xi}}_{-1} = \frac{1}{\sqrt{2}} (1, {\it -i}, 0)~~~~~~~ 
{\vec{\xi}}_0 = (0, 0, 1), 
\end{eqnarray} 
${\bf p}$ is the relative 
proton--neutron momentum inside the deuteron, 
$\hat{\bf p}={\bf p}/{|{\bf{p}}|}$ is the unit vector in the ${\bf p}$ 
direction.  Here $S$ and $D$ components are defined as 
\begin{eqnarray}
\label{C-dwf}
U(p)&=&u(p) + v_o(p)\cdot e^{{\it i}\chi},\nonumber\\
W(p)&=&w(p) + v_2(p)\cdot e^{{\it i}\chi},
\end{eqnarray}
where
$u(p)$  and $w(p)$ are $S$ and $D$ components of
the standard deuteron wave function
based on the $NN$ potentials; $v_o(p)\cdot e^{{\it i}\chi}$ 
and $v_2(p)\cdot e^{{\it i}\chi}$
are the complex $6q$ admixtures to the $S$ and $D$ components of
the standard DWF, respectively.

Using parity conservation, time reversal invariance and the Pauli principle
we can write the  matrix of $NN$ elastic scattering in terms of 5  
independent complex amplitudes \cite{lehar} (when isospin invariance is 
assumed):  
\begin{eqnarray} 
\label{NNmatrix} 
M({\bf k}^\prime,{\bf k})=\frac{1}{2}
& &\big ((a+b)+ (a-b) (\vec{\sigma}_1\cdot{\bf{n}})\cdot 
(\vec{\sigma}_2\cdot{\bf{n}})+ \nonumber\\ +& &(c+d) 
(\vec{\sigma}_1\cdot{\bf{m}})\cdot 
(\vec{\sigma}_2\cdot{\bf{m}})+\nonumber\\ 
+& & (c-d) 
(\vec{\sigma}_1\cdot{\bf{l}})\cdot (\vec{\sigma}_2\cdot{\bf{l}})+ 
e((\vec{\sigma}_1+\vec{\sigma}_2)\cdot{\bf{n}})\big ),
\end{eqnarray}
where $a$, $b$, $c$, $d$ and $e$ are the scattering amplitudes,
$\vec{\sigma}_1$ and $\vec{\sigma}_2$ are the Pauli $2\times 2$ matrices,
{\bf{k}} and {\bf{k}}$^\prime$ are the unit vector in the direction of 
the incident and scattered particles, respectively, and center-of-mass 
basis vectors {\bf{n}}, {\bf{m}}, {\bf{l}} are defined as:  
\begin{eqnarray}
{\bf{n}}=\frac{{\bf{k}^\prime}\times {\bf{k}}}
{|{\bf{k}^\prime}\times 
{\bf {k}}|},~~~~~~ 
{\bf{l}}=\frac{\bf{\vec{k}^\prime}+{\bf{k}}}{|{\bf{k}^\prime}+
{\bf{k}}|},~~~~~~
{\bf{m}}=\frac{{\bf{k}^\prime}-{\bf{k}}}
{|{\bf{k}^\prime}-{\bf{k}}|}.~~~~~~
\end{eqnarray}
However, at a zero angle there are only 3 independent amplitudes and 
the matrix element (\ref{NNmatrix}) 
can be written as \cite{lehar}
\begin{eqnarray}
\label{NN0}
{\cal M}(0)= \frac{1}{2} \big ( A + B 
(\vec{\sigma}_1\cdot\vec{\sigma}_2) +C 
(\vec{\sigma}_1\cdot{\bf{k}})(\vec{\sigma}_2\cdot{\bf{k}})\big ), 
\end{eqnarray} 
where amplitudes $A$, $B$ and $C$ are related to the 
amplitudes defined in ref.\cite{lehar} as follows 
\begin{eqnarray} 
A= a(0) + b(0),~~~~~~B= c(0)+ d(0),~~~~~~C= -2 d(0).
\end{eqnarray}

We consider  deuteron breakup reaction in the special kinematics, i.e.
with the emission of the spectator proton at zero angle, while the neutron
interacts with the proton target and the products of this interaction go along
the axis of the reaction.

Using of both  (\ref{dwf}) and  (\ref{NN0}) expressions
the matrix element of deuteron breakup process in collinear geometry
can be written as \cite{lad97}
\begin{eqnarray}
\label{m_break}
{\cal M}=& &\frac{\it i}{2 \sqrt{2}} \frac {1}{\sqrt{4\pi}}
\psi^{+}_{1}
\psi^{+}_{2}
\left [\left ( U(p) (\vec{\sigma}\cdot\vec{\xi}) -\frac{W(p)}{\sqrt{2}}
(3 
(\hat{\bf 
p}\vec{\xi})(\vec{\sigma}\hat{\bf 
p})-(\vec{\sigma}\cdot\vec{\xi}))\right ) \sigma_y \right 
]\times \nonumber\\ 
\times & & \big ( A + B (\vec{\sigma}_1,\vec{\sigma}_2) +C 
(\vec{\sigma}_1,{\bf{k}})(\vec{\sigma}_2,{\bf{k}})\big ) \psi^{*}_{1} 
\psi_{2}.
\end{eqnarray}

The matrix 
element of $dp$ backward elastic scattering 
within framework of one nucleon exchange  
has the following form   
\begin{eqnarray} 
\label{m_dpback}
{\cal M} = \frac{1}{8\pi}\psi^{ +}_{f} & & \left ( U^*(p) 
(\vec{\sigma_f}\cdot\vec{\xi}^*_f) -\frac{W^*(p)}{\sqrt{2}} (3 
(\hat{\bf p}\vec{\xi}^*_f)(\vec{\sigma_f}\hat{\bf p})-
(\vec{\sigma_f}\cdot \vec{\xi}^*_f))\right )\times
\nonumber\\ \times& & \left (U(p) 
(\vec{\sigma_i}\cdot\vec{\xi}_i) -\frac{W(p)}{\sqrt{2}} (3 
(\hat{\bf p}\vec{\xi}_i)(\vec{\sigma_i}\hat{\bf p})-
(\vec{\sigma_i}\cdot\vec{\xi}_i))\right )\psi_{i} 
\end{eqnarray} 

\section{Six-quark configurations}

In this section we consider different models considering
the 
quark (or baryon--baryon) degrees of freedom inside the deuteron.

In the hybrid model of the DWF \cite{kob1} 
the  $6q$ amplitude, arising from the $s^6$ configurations of
six quarks, must be added to the  $S$ component of the
standard DWF according the following expression 
\begin{eqnarray}
U(k)=\sqrt{1-\beta^2}\cdot u(k) + \beta\cdot v_0(k) \cdot e^{i\chi},
\end{eqnarray}
where parameter $\beta$ and phase $\chi$  represent 
value of $6q$ admixture in deuteron and the degree of 
non--orthogonality between $np$ and $6q$ components of the DWF, 
respectively.

The $6q$ admixture has the following form
\begin{eqnarray}
v_0(k) = I\sqrt{10}\cdot 2^{2/3}\left (\frac{2}{1+\sqrt{2}}\right )^{6} 
\left (\frac{2}{3\pi\omega}\right )^{3/4} e^{-k^2/3\omega}
\end{eqnarray}
Factor $I\approx 0.332$ is the overlap factor of color
spin--isospin wave fuctions and $\omega$ defines the root-mean-square 
radius of the $6q$ configuration $r^2 =5/4\omega$,
$k$ is  
internal momentum of a nucleon
in the deuteron defined in the light-cone dynamics
\cite{dirac}-\cite{fs1}.

The parameters of the $6q$ admixture $r$, $\beta$ and $\chi$ were
obtained in \cite{ladphd} from the fit of 
the experimental data on the 
momentum density of the nucleon in deuteron $\phi^2(k)$ \cite{deucr},
tensor analyzing power $T_{20}$ \cite{ladygin}
and polarization transfer
coefficient $\kappa_0$ \cite{kappa0,kappa1,kappa2}  
for deuteron inclusive breakup 
reaction with the emission of the proton at a zero angle 
using standard DWFs \cite{paris,rsc,bonn}.
The results of the fit are given in Table 1 and shown
in Fig.1 by the solid, dashed and dotted lines for 
RSC \cite{rsc}, Paris \cite{paris} and Bonn (version C) \cite{bonn}
DWFs, respectively. One can see the satisfactory description of the 
experimental data. The probability  of the $6q$ admixture is found
to be $3-4\%$. The relative phase is $\sim 40^0$ for 
RSC \cite{rsc} and Paris \cite{paris} DWFs and $55^0$ for
Bonn C DWF \cite{bonn}. The 
radius is $r\sim 0.6$ fm for all used DWFs.
The parameters are comparable with 
the results obtained in \cite{t20br1} using RSC DWF \cite{rsc}.

In \cite{lykasov,efremov} the nonnucleon degrees of freedom
($NN^*$, $NN\pi$ and higher components of the Fock space)
were taken into account in the following way
\begin{eqnarray}
\Phi^2(\alpha,k_t) = (1-\beta^2) \phi_{NN}^2(\alpha,k_t)/
(2\alpha(1-\alpha)) +\beta^2 G_d(\alpha,k_t),
\end{eqnarray}
where $\Phi^2(\alpha,k_t)$ is the distribution of constituents in
the deuteron; \\
$\phi_{NN}^2(\alpha,k_t)$ is the relativized standard 
DWF; $G_d(\alpha,k_t)$ is the distribution of $NN^*$, $NN\pi$ ..., or 
$6q$ component in the deuteron. Parameter $\beta^2$ gives the 
probability of this nonnucleon component.
The relativistic form of the DWF $\phi_{NN}(\alpha,k_t)$ can be written 
according \cite{dirac}-\cite{fs1} as 
\begin{eqnarray} 
\phi_{NN}(\alpha,k_t)=\left ( \frac{m_p^2+k_t^2}{4\alpha(1-\alpha)}
\right )^{1/4} \phi(k),
\end{eqnarray}
where $\phi(k)$ is the standard DWF (for 
instance, \cite{paris,rsc,bonn}) and  internal momentum $k$ and 
longitudinal momentum fraction $\alpha$ are defined as 
\cite{dirac}-\cite{fs1} 
\begin{eqnarray} 
k^2&=&\frac{m_p^2+k_t^2}{4\alpha(1-\alpha)} - m_p^2 ,
\nonumber\\
\alpha &=& \frac{k_{||}+\sqrt{k_{||}^2+m_p^2}}{2\sqrt{k_{||}^2+m_p^2}}.
\end{eqnarray}
Here $k_{||}$ is the longitudinal momentum in infinite momentum
frame and $m_p$ is the nucleon mass.

The expression for nonnucleon component $G_d(\alpha,k_t)$ 
is written as \cite{efremov}
\begin{eqnarray}
\label{gd}
G_d(\alpha,k_t) = b^2/(2\pi)G_1(\alpha) e^{-bk_t}
\end{eqnarray}
with 
\begin{eqnarray}
\label{g1}
G_1(\alpha) = \Gamma(A_2+B_2+2)/(\Gamma(A_2+1)\Gamma(B_2+1))
\alpha^{A_2} (1-\alpha)^{B_2},
\end{eqnarray}
where $\Gamma(...)$ denotes the $\Gamma$-function. 
The parameter $b$ is chosen to be 5 GeV/c.
We assume, that nonnucleon component (\ref{gd})-(\ref{g1}) has the 
relative phase $\chi$ with the $S$ wave of the standard DWF 
\cite{lykasov}.

The results of the fit of the experimental data
\cite{deucr,ladygin,kappa0,kappa1,kappa2}  
are given in Table 2 and 
shown in Fig.2 by the solid, dashed and dotted lines for RSC 
\cite{rsc}, Paris \cite{paris} and Bonn (version C) \cite{bonn} DWFs, 
respectively. 
The probability  of the nonnucleon 
component is found
to be also $\sim 3\%$. The relative phase $\chi$ between
$NN$ and nonnucleon components is $~40-60^o$. 
The parameters $A_2$ and $B_2$ are  found to be 
approximately the same for Paris \cite{paris}, RSC \cite{rsc}
and Bonn C \cite{bonn} DWFs.
Note, 
all the used $NN$ DWFs provide satisfactory  agreement
with the existing data, however,
the using of
RSC DWF gives  better description of the polarization
transfer coefficient $\kappa_0$.

\section{T-odd polarization effects}

Let us define the general spin observable of the third order in
terms of Pauli $2\times 2$ spin matrices $\sigma$ for protons and a set of
spin operators $S_{\lambda}$ for the spin 1 particle for both reactions 
as \cite{argonne}
\begin{eqnarray}
\label{general}
C_{\alpha,\lambda,\beta,0}=
\frac{Tr({\cal M}{\sigma^p_\alpha}{S^d_\lambda}{\cal M^+}{\sigma^p_\beta})}{Tr({\cal M}{\cal M^+})},
\end{eqnarray}
where indices $\alpha$ and $\lambda$ refer to the initial proton and
deuteron polarization, index $\beta$ refers to the final  proton, 
respectively.

We use a righthand coordinate system,
defined in accordance with Madison convention \cite{madison}.
This system is specified by a
set of three orthogonal vectors $\vec{L}$, $\vec{N}$ and  $\vec{S}$,
where $\vec{L}$ is the unit vector along the 
momenta  of the incident particle,
$\vec{N}$
is taken to be orthogonal to $\vec{L}$,
$\vec{S}=\vec{N}\times\vec{L}$.

In this paper we consider $T$-odd polarization observables,
namely: tensor- vector spin correlations $C_{N,SL,0,0}$ due to
tensor polarization of the beam and polarization of the initial
proton and polarization transfer coefficient $C_{0,SL,N,0}$ from
tensor polarized deuteron to proton in the $dp\to pd$ and 
$dp\to p(0^0)+p(180^0)+n$ reactions. 
Note, that such observables must be zero  
in framework of one nucleon exchange using standard 
deuteron wave 
functions, however, they do not vanish with the existing of 
$6q$ admixture in the DWF.

Using the formulas for
the matrix elements of the $p(d,p)pn$
and $p(d,p)d$ reactions (\ref{m_break}) and 
(\ref{m_dpback}), respectively, one can obtain
the expression for the polarization transfer coefficient
$C_{0,SL,N,0}$ 
\begin{eqnarray} 
\label{trans} 
C_{0,SL,N,0}= 
\frac{3}{\sqrt{2}}\frac{w v_0 sin\chi} {u^2+w^2+v_0^2+2 u v_0 
cos\chi}. 
\end{eqnarray}
One can see that $C_{0,SL,N,0}$ does not depend on the initial
energy and is defined only by the interference between $D$ wave
of the standard DWF and $6q$ admixture. The results of the 
calculations with the use of
Paris \cite{paris}, RSC \cite{rsc}
and Bonn C \cite{bonn} DWFs are presented in Fig. 3 {\it a}, 
{\it b} and {\it c} 
for two different models of the $6q$ admixture: \cite{kob1} and 
\cite{lykasov,efremov} given by the solid and dashed lines, 
respectively.  These two types of the hybrid DWFs give quite similar 
behaviour of the $C_{0,SL,N,0}$ up to $k\sim 800$ MeV/c,
however, they differ at higher momenta.            
Both models predict the smooth variation of the $C_{0,SL,N,0}$
of about $-1$ at $k\sim 600$ MeV/c.
The dependence on the used $NN$ 
deuteron wave function occurs only at high $k$ of about $900$ MeV/c, 
therefore, the observation of large negative value of $C_{0,SL,N,0}$ 
could indicate that quark degrees of freedom play quite important role 
in the deuteron at large $k$.

Spin correlation parameter $C_{N,SL,0,0}$ due to tensor
polarization of the beam and polarization of the initial
proton for the $dp\to p(0^0)+p(180^0)+n$ process can be written 
as
\begin{eqnarray} 
\label{spdpppn} 
C_{N,SL,0,0}= 
\frac{3}{\sqrt{2}}\frac{w v_0 sin\chi} 
{u^2+w^2+v_0^2+2 u v_0 cos\chi} 
\cdot A_{oonn}(0^o),
\end{eqnarray}
where $A_{oonn}(0^o)$ is spin correlation of
neutron--proton elastic scattering at a zero angle
for vertically polarized particles (see notations used
in \cite{lehar,lad97}).
Therefore, the behaviour of $C_{N,SL,0,0}$ 
in the $dp\to p(0^0)+p(180^0)+n$ reaction
is defined both the DWF and
$np$ elementary amplitude which is energy dependent. 
The calculation of $C_{N,SL,0,0}$ 
for the deuteron initial energy of 2.1 GeV and
1.25 GeV using the results
of phase-shift analysis performed
in \cite{vpi} are shown in Figs 4 and 5, 
respectively.  One can see that $C_{N,SL,0,0}$ is positive at 2.1 GeV 
up to $k\sim 550$ MeV/c and negative at 1.25 GeV at $k\sim 
300\div 400$ MeV/c. 
The difference between two models of $6q$ admixture
shown by the solid \cite{kob1} and dashed  
\cite{lykasov,efremov} lines in {\it a}, {\it b} and  {\it c}
figures for 
Paris \cite{paris}, RSC \cite{rsc}
and Bonn C \cite{bonn} DWFs 
is not dramatic at both energies.

Spin correlation parameter 
$C_{N,SL,0,0}$ in deuteron--proton backward elastic scattering
is given in the following form 
\begin{eqnarray} 
\label{spdppd} 
C_{N,SL,0,0}= 
\frac{1}{\sqrt{2}}\frac{w v_0 sin\chi
\cdot \left ((u +v_0 cos\chi - \sqrt{2} w)^2 +v_0^2 sin^2\chi\right ) } 
{(u^2+w^2+v_0^2+2 u v_0 cos\chi)^2}. 
\end{eqnarray}
The behaviour of this observable for different
types of $6q$ admixture in the DWFs is shown in Fig. 6 {\it a},
{\it b} and {\it c} for
Paris \cite{paris}, RSC \cite{rsc}
and Bonn C \cite{bonn} DWFs 
by the solid \cite{kob1} and dashed
\cite{lykasov,efremov} lines, respectively.
One can see  that 
the spin correlation $C_{N,SL,0,0}$ 
has a small negative
value at low $k$, then it approaches a minima
of $\sim -0.7\div -0.8 $ at $k\sim 
400$ MeV/c and afterwards it goes smoothly to a zero
for both models of $6q$ component.  However,
the use of DWFs with the $6q$ admixture adopted in 
\cite{lykasov,efremov} gives systematically more negative value  
of spin correlation at 
internal momenta of $\ge 300$ MeV/c. 
The use of different $NN$ potentials \cite{paris,rsc,bonn}
(see Fig.6 {\it a}, {\it b} and  {\it c}, respectively)
gives  slightly different behaviour of $C_{N,SL,0,0}$ 
for both models of $6q$ 
admixture. Nevertheless,  one can conclude that the measurements of 
spin correlation $C_{N,SL,0,0}$ in $dp$ backward elastic scattering can 
help to distinguish between these two models.
 
Note that non--orthogonality in the deuteron wave function results
in the $T$-invariance violation, which contradicts the
experiment.   However, $NN$ and $6q$ components can be
orthogonalized following the procedure described in 
ref.\cite{burov}. Such a procedure only slightly changes the
probability of $6q$ admixture 
\cite{burov}, but does not affect on the behaviour
of the considered observables. For instance,  the probability
of $6q$ component changes from $2.96\%$  to $3.31\%$ and
from $3.42\%$ to $4.17\%$ for the models  
\cite{kob1} and 
\cite{lykasov,efremov}, respectively, in the case of the use
of Paris DWF \cite{paris}.

The six-quark wave function of the 
deuteron  has been calculated recently not only from $s^6$ , but also 
from $s^4p^2$ configurations \cite{obukh}.
Such  configurations are orthogonal to the 
usual $S$ and $D$ waves in the deuteron. 
Tensor analyzing power $T_{20}$ and polarization transfer coefficient
$\kappa_0$ in deuteron inclusive breakup at a zero proton emission angle
have been qualitatively reproduced 
at large internal momenta using the results of these calculations
\cite{gorovoj}. 
The probability of the $D$ wave 
originated from $s^4p^2$ configurations was found to be about $0.5\%$
(a small part of $D$ wave probability $\sim 6\%$).
The results on the polarization transfer $C_{0,SL,N,0}$ and 
spin correlation $C_{N,SL,0,0}$ in $dp$ backward elastic scattering 
using Paris DWF \cite{paris} and the $6q$ projection on
$NN$ component from \cite{gorovoj} are given in 
Fig.7 {\it a} and {\it b}, respectively. 
The behaviour of these observable differ significantly 
from the results
shown in Figs. 3 and 6. This deviation is due to 
presence of $D$- wave in six-quark wave function.
The results on tensor- vector
spin correlation $C_{SL,N,0,0}$ in the reaction
$dp\to p(0^0)+p(180^0)+n$  at the initial deuteron energy
of 2.1 and 1.25 GeV are shown in Fig.8 {\it a} and {\it b}, 
respectively.  The behaviour                 is qualitatively the same 
as shown in  Figs. 4 and 5, however, the value of $C_{SL,N,0,0}$ at 2.1 
GeV and $k\sim 300$ MeV/c is as twice as much than that in the case of 
absence of $D$ wave.   

Note, the one 
of interesting features of QCD is the possible existing of 
resonances in the dibaryon system  corresponding
to six-quark states which are
dominantly hidden color, i.e., orthogonal to the usual $np$ states.
The rich structure in the behaviour of the tensor analyzing
power $T_{20}$ in $dp$ backward elastic scattering  
\cite{dppddub,dpback249} can be an
indication of such dibaryon resonances \cite{prop}.

Of course, the mechanisms additional to ONE can contribute to
$C_{N,SL,0,0}$ and $C_{0,SL,N,0}$.
However, the calculations taking 
into account such mechanisms \cite{lykasov} show that their
 contribution is small at large internal momenta.
Thus, the observation of a large values of 
$C_{N,SL,0,0}$ and $C_{0,SL,N,0}$ at momenta higher
600 MeV/c could be rather clear indication of the 
exotic $6q$ configurations.

%\newpage

\section{Conclusion}

We have considered $T$-odd observables
in deuteron 
exclusive breakup and $dp$ 
backward elastic scattering, namely,
tensor- vector polarization transfer coefficient
$C_{0,SL,N,0}$ and  
tensor- vector spin correlation $C_{N,SL,0,0}$.
These observables, which are  associated with the tensor
polarization of the deuteron and polarization of the proton,
show their sensitivity to the quark degrees of freedom in the
deuteron and their spin structure. 
The calculations give a sizeable effects at 
large internal momenta, which could be measured with the existing 
experimental techniques.

Measurements of these observables could be performed at COSY at Zero
Degree Facility (ANKE) using internal polarized target with the
detection of two charged particles in case of deuteron breakup and
with the detection of the fast proton in case of $dp$ backward elastic
scattering.

Such experiments could be also performed at the Laboratory for High 
Energies of  Joint Institute 
for Nuclear Researches.
The rotation of the primary deuteron
spin could be provided by the magnetic
field of the beam line upstream of the target or by
the special spin-rotating magnet.

\vspace{0.3cm}

\begin{center}
{\Large\bf Acknowledgments}
\end{center}

\begin{sloppypar}
I'm indebted to Dr. N.B.Ladygina for critically
revising of this manuscript and for permanent 
help and support.
Author is grateful to Prof. A.P.Kobushkin
for helpful comments and discussions.
I also thank I.M.Sitnik for estimation of the bending
angle of the magnetic system upstream of the polarized target
installed now at the LHE of JINR.
\end{sloppypar}

%\newpage

%\vspace{1cm}

\newpage

Table 1. Parameters of the $6q$ admixture in the hybrid
model \cite{kob1} for different standard DWFs \cite{paris,rsc,bonn}.

\begin{center}
\begin{tabular}{|c|c|c|c|}
\hline
DWF & $\beta^2,~\%$ & $\chi$ & $r,~fm$\\
\hline
\cite{paris} & $3.42\pm 0.09$ & $47.2^0\pm 0.6^0$ & $0.578\pm 0.009$\\
\cite{rsc}   & $4.07\pm 0.10$ & $40.1^0\pm 0.6^0$ & $0.590\pm 0.009$\\
\cite{bonn}  & $2.79\pm 0.09$ & $55.1^0\pm 0.7^0$ & $0.595\pm 0.010$\\
\hline
\end{tabular}
\end{center}

Table 2. Parameters of the $6q$ admixture 
\cite{lykasov,efremov} for different standard DWFs \\
\cite{paris,rsc,bonn}.

\begin{center}
\begin{tabular}{|c|c|c|c|c|}
\hline
DWF & $\beta^2,~\%$ & $A_2$ & $B_2$ & $\chi$ \\
\hline
\cite{paris} & $2.96\pm 0.18$ & $10.0^*~~~~~~~$& $20.23\pm 0.42$ &
$47.0^0\pm 0.6^0$ \\
\cite{rsc}   & $3.70\pm 0.43$ & $10.0\pm 0.9$ & $19.87\pm 2.72$ &
$40.2^0\pm 0.6^0$ \\ 
\cite{bonn}  & $2.67\pm 0.19$ & $10.0^*~~~~~~~$ & $19.46\pm 0.48$ &
$54.6^0\pm 0.7^0$ \\ 
\hline 
\end{tabular} 
\end{center}
$*$ - parameter is fixed.

\newpage
\begin{center}
{\Large Figure captions}
\end{center}

Fig.1. Momentum density $\Phi^2(k)$ \cite{deucr},
tensor analyzing power $T_{20}$ \cite{ladygin} and
polarization transfer coefficient $\kappa_0$ 
\cite{kappa0}
(open squares),  \cite{kappa1} (full triangles) and
\cite{kappa2} (full circles and squares) 
versus internal momentum $k$ in 
deuteron inclusive breakup with the emission of proton at $0^0$.
Full, dashed and dotted lines correspond to
calculations with hybrid wave function \cite{kob1} using 
RSC \cite{rsc}, Paris \cite{paris} and Bonn C \cite{bonn}
DWFs, respectively.
\vspace{0.3cm}

Fig.2. Momentum density $\Phi^2(k)$ \cite{deucr},
tensor analyzing power $T_{20}$ \cite{ladygin} and
polarization transfer coefficient $\kappa_0$ 
\cite{kappa0,kappa1,kappa2}
versus internal momentum $k$ in 
deuteron inclusive breakup with the emission of proton at $0^0$.
Full, dashed and dotted lines correspond to
calculations with the wave function adopted in \cite{lykasov,efremov} 
using RSC \cite{rsc}, Paris \cite{paris} and Bonn C \cite{bonn} DWFs, 
respectively.
The symbols are the same as in Fig.1.
\vspace{0.3cm}

Fig.3. Tensor-vector polarization transfer coefficient 
$C_{0,SL,N,0}$ in deuteron exclusive breakup 
in the collinear geometry and $dp$
backward elastic scattering using $6q$ admixture adopted in
\cite{kob1}               and \cite{lykasov,efremov}
and given by the solid and dashed lines,
respectively. The curves in Figs. {\it a}, {\it b} and {\it c} are
obtained with the 
use of                 Paris \cite{paris}, RSC \cite{rsc}
and Bonn C \cite{bonn} DWFs, respectively. 
\vspace{0.3cm}

Fig.4. Tensor-vector spin correlation parameter 
$C_{N,SL,0,0}$ in deuteron exclusive breakup 
in the collinear geometry  at 2.1 GeV of the deuteron initial energy  
using $6q$ admixture adopted in
\cite{kob1} (solid lines) and \cite{lykasov,efremov}(dashed lines).
The curves in Figs. {\it a}, {\it b} and {\it c} are   
obtained with the 
use of                 Paris \cite{paris}, RSC \cite{rsc}
and Bonn C \cite{bonn} DWFs, respectively. 
\vspace{0.3cm}

Fig.5. Tensor-vector spin correlation parameter 
$C_{N,SL,0,0}$ in deuteron exclusive breakup 
in the collinear geometry  at 1.25 GeV of the deuteron initial energy  
using $6q$ admixture adopted in
\cite{kob1} and \cite{lykasov,efremov}
and given by the solid and dashed lines,
respectively. 
The curves in Figs. {\it a}, {\it b} and {\it c} are   
obtained with the 
use of                 Paris \cite{paris}, RSC \cite{rsc}
and Bonn C \cite{bonn} DWFs, respectively. 
\vspace{0.3cm}

Fig.6. 
Tensor-vector spin correlation parameter 
$C_{N,SL,0,0}$ in deuteron- proton
backward elastic scattering 
using $6q$ admixture adopted in
\cite{kob1} and \cite{lykasov,efremov}
and given by the solid and dashed lines,
respectively. 
The curves in Figs. {\it a}, {\it b} and {\it c} are   
obtained with the 
use of                 Paris \cite{paris}, RSC \cite{rsc}
and Bonn C \cite{bonn} DWFs, respectively. 
\vspace{0.3cm}

Fig.7. 
a) Tensor-vector polarization transfer coefficient 
$C_{0,SL,N,0}$ in deuteron exclusive breakup 
in the collinear geometry and $dp$
backward elastic scattering and 
b) 
tensor-vector spin correlation parameter 
$C_{N,SL,0,0}$ in deuteron--proton
backward elastic scattering
using results of 
\cite{gorovoj} and Paris DWF \cite{paris}.
\vspace{0.3cm}

Fig.8. Tensor- vector spin correlation parameter 
$C_{N,SL,0,0}$ in deuteron exclusive breakup 
in the collinear geometry  at a) 
2.1 GeV and b) at 1.25 GeV of the deuteron initial 
energy results of
\cite{gorovoj} and Paris DWF \cite{paris}.

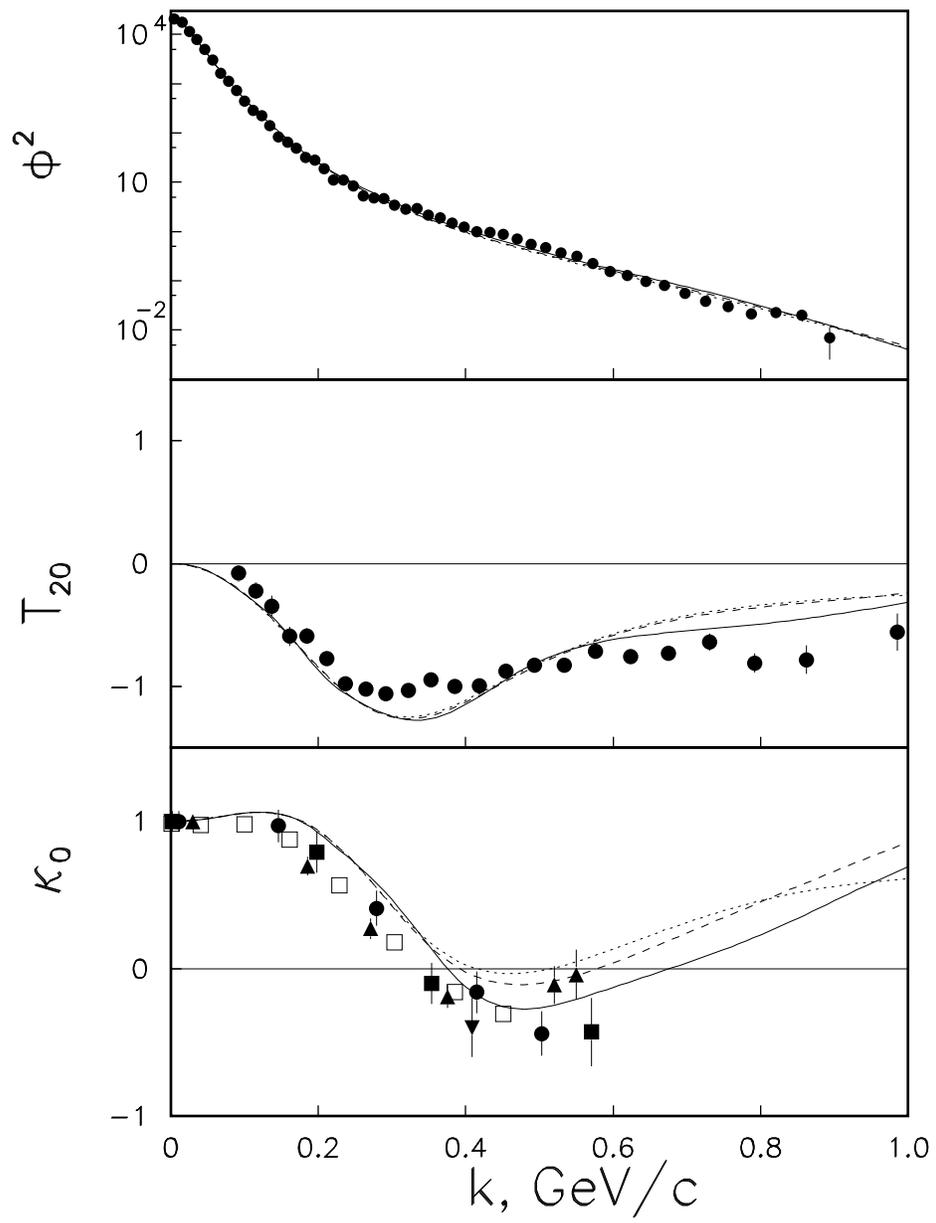
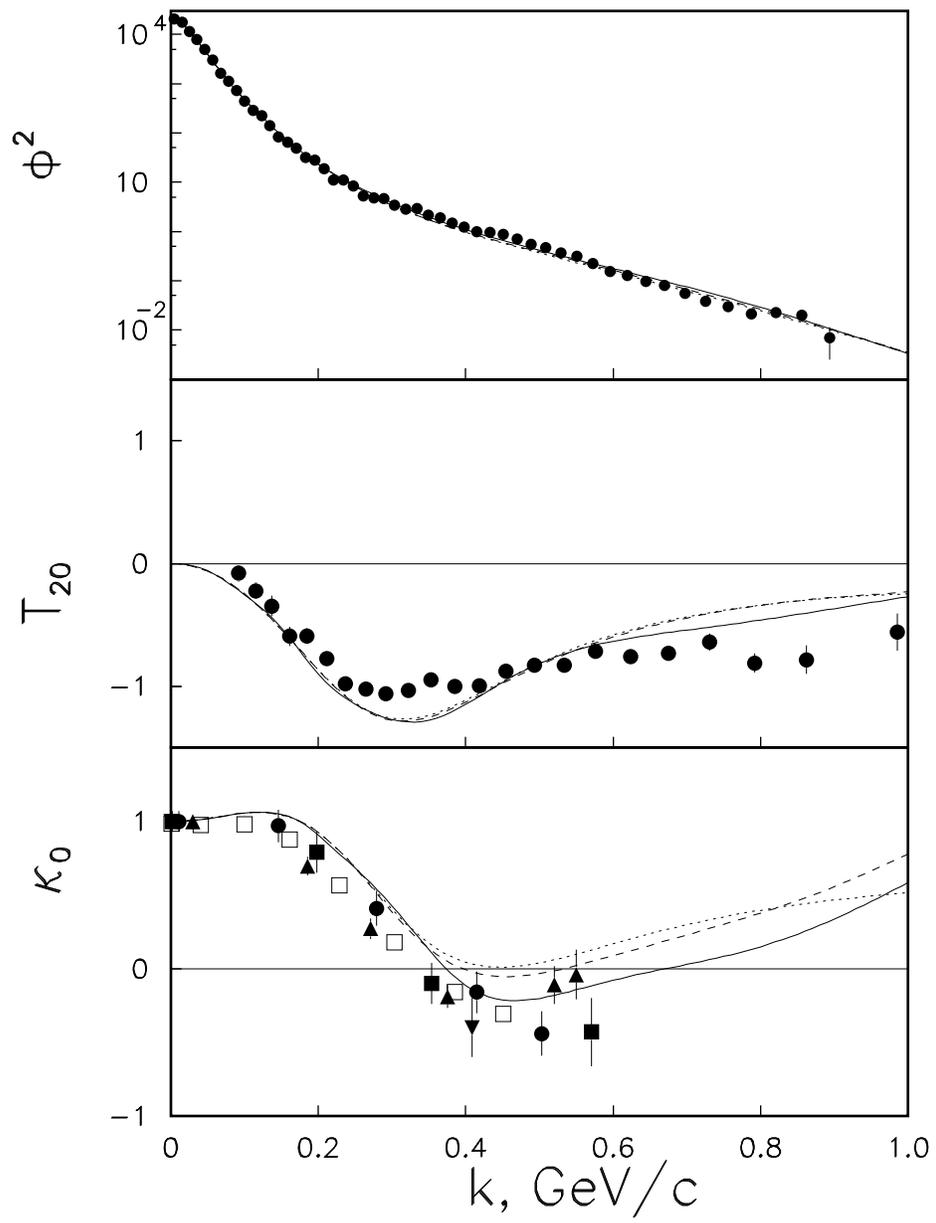
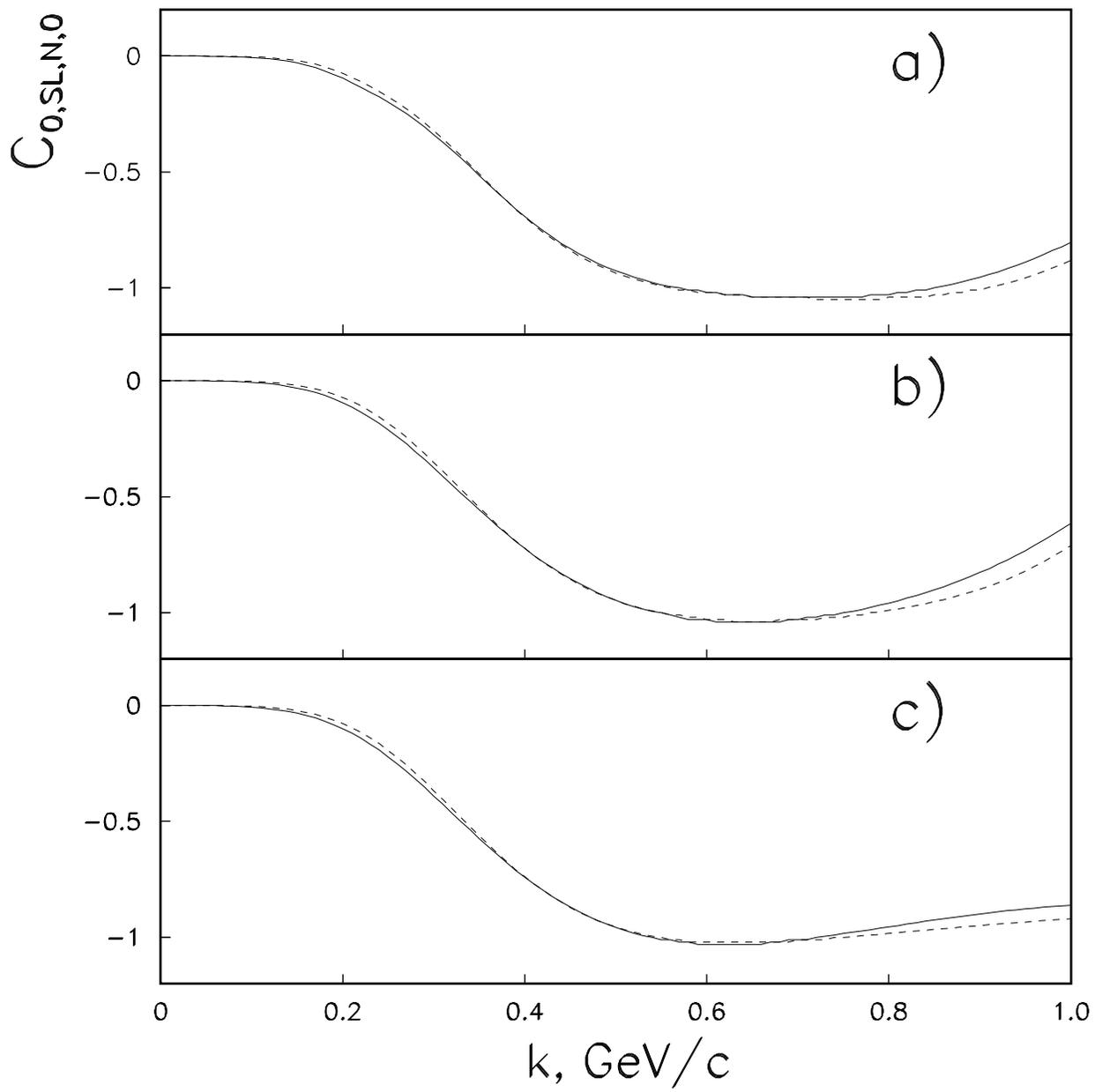
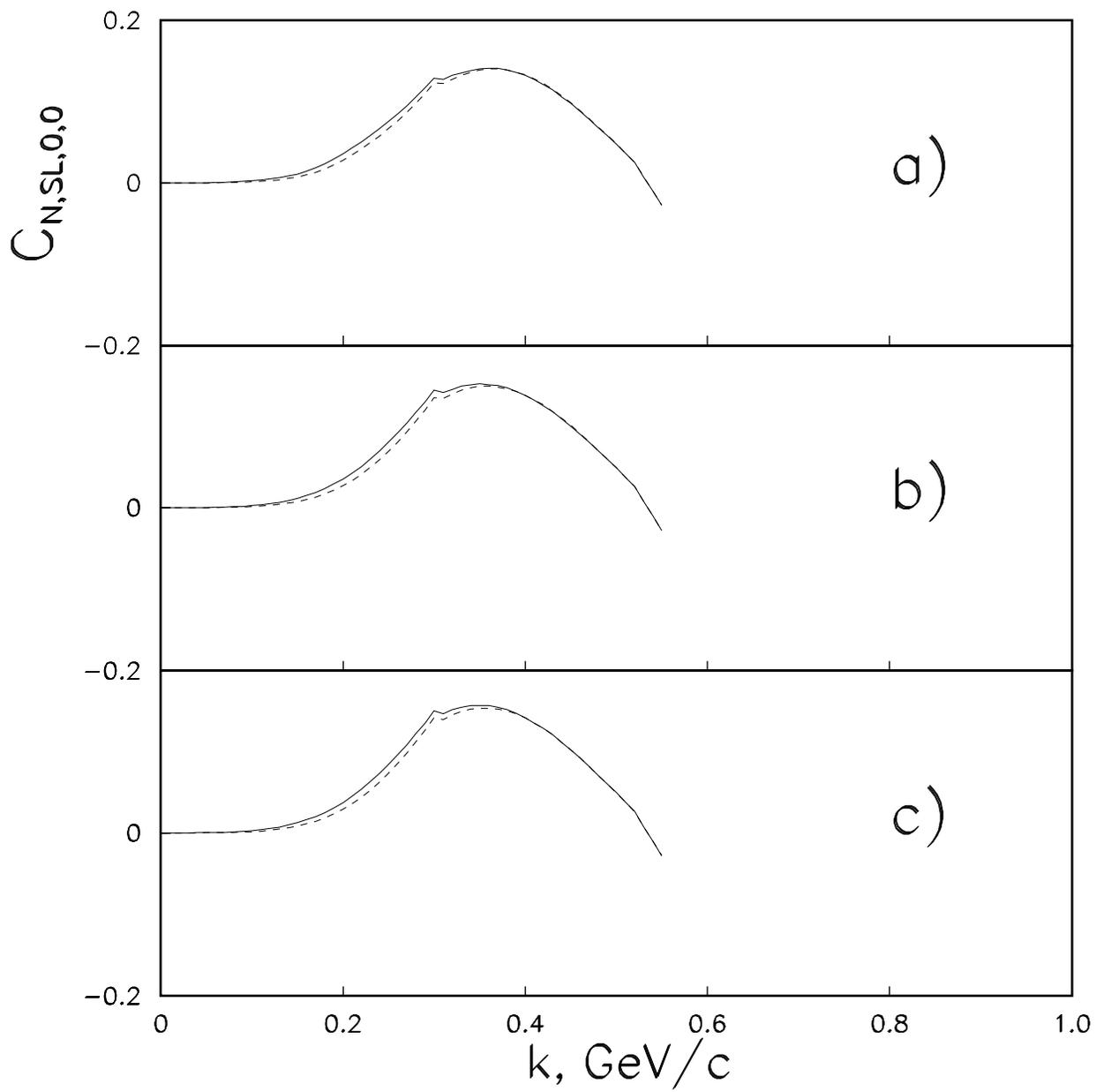
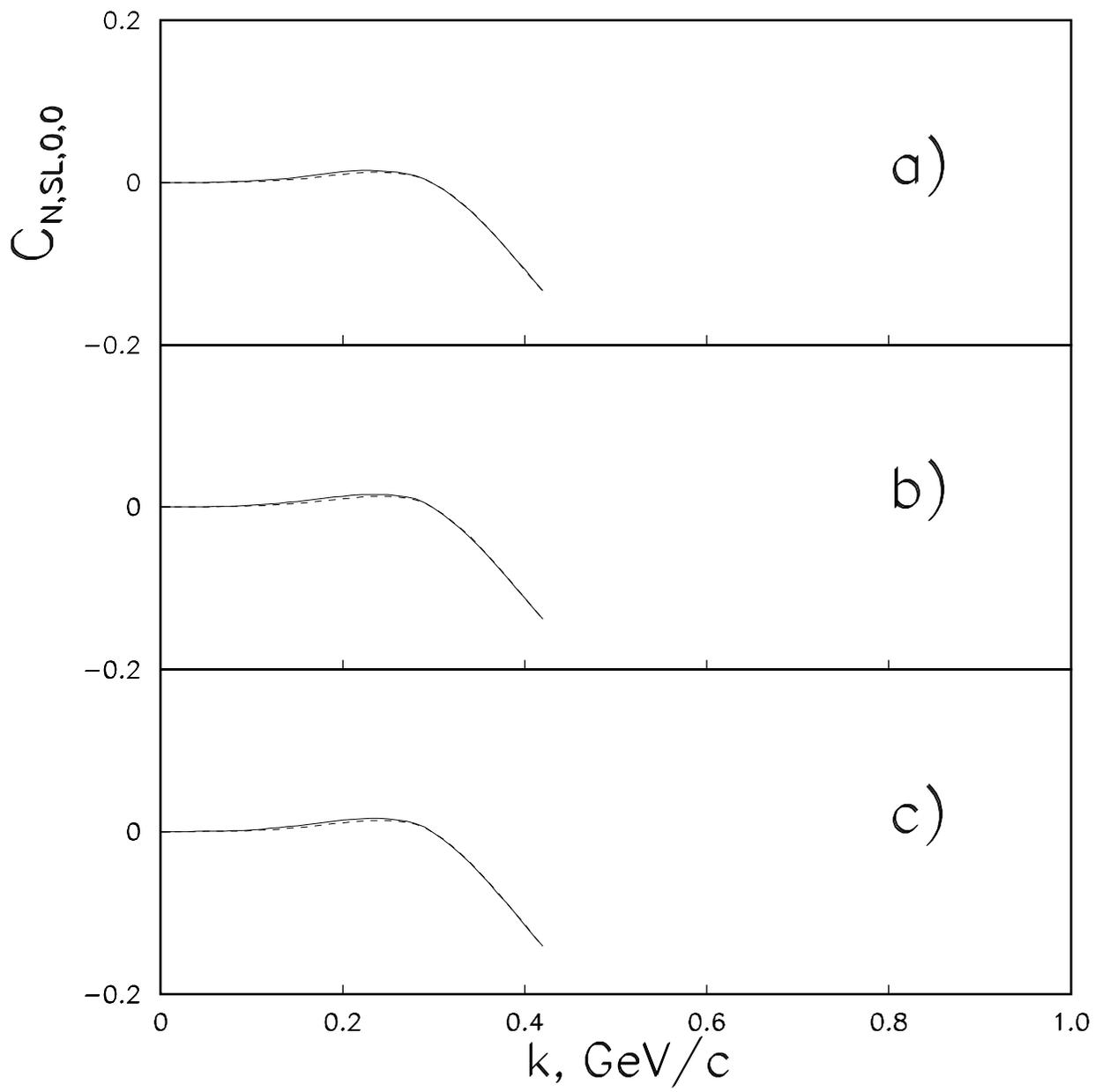
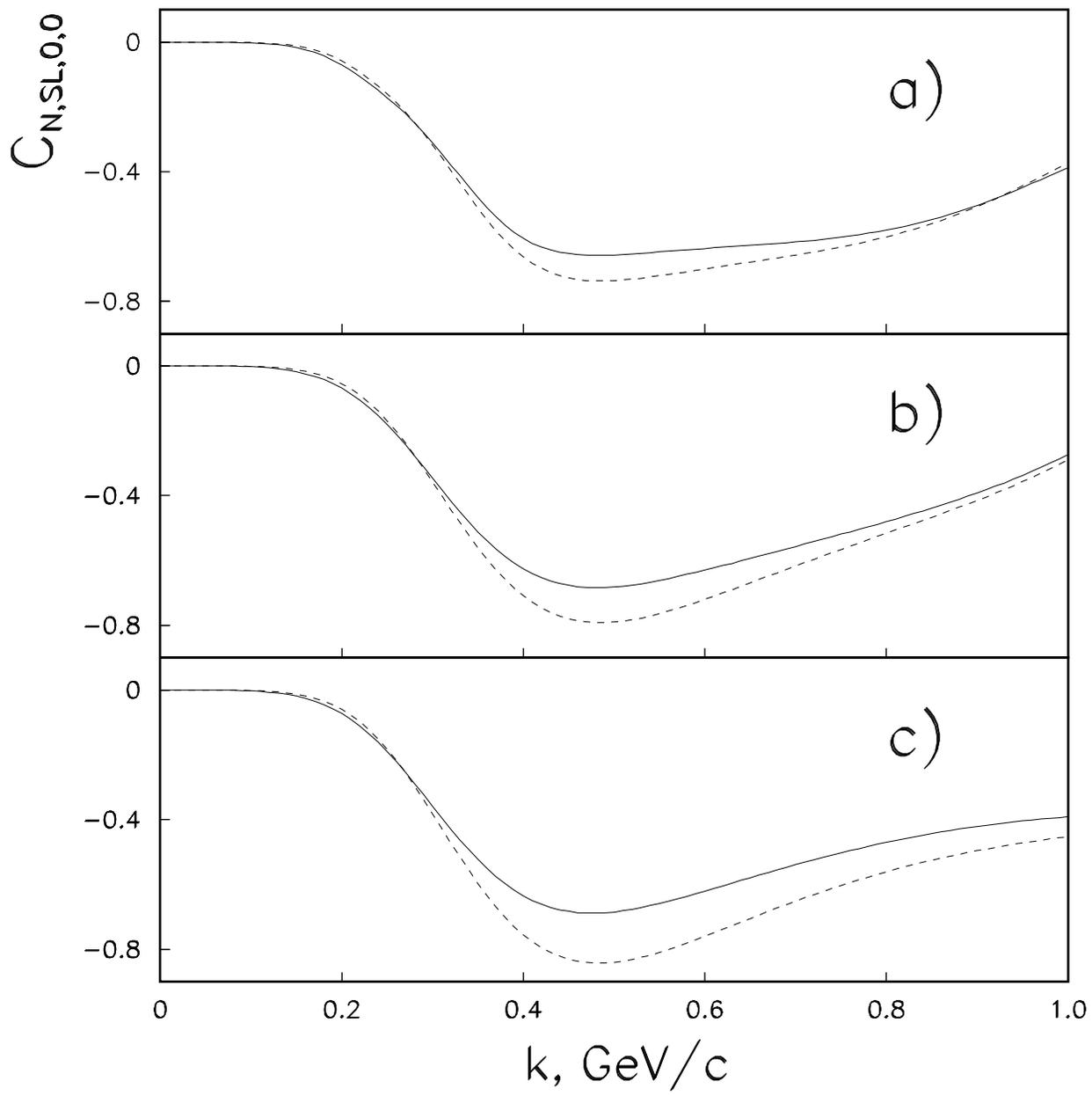
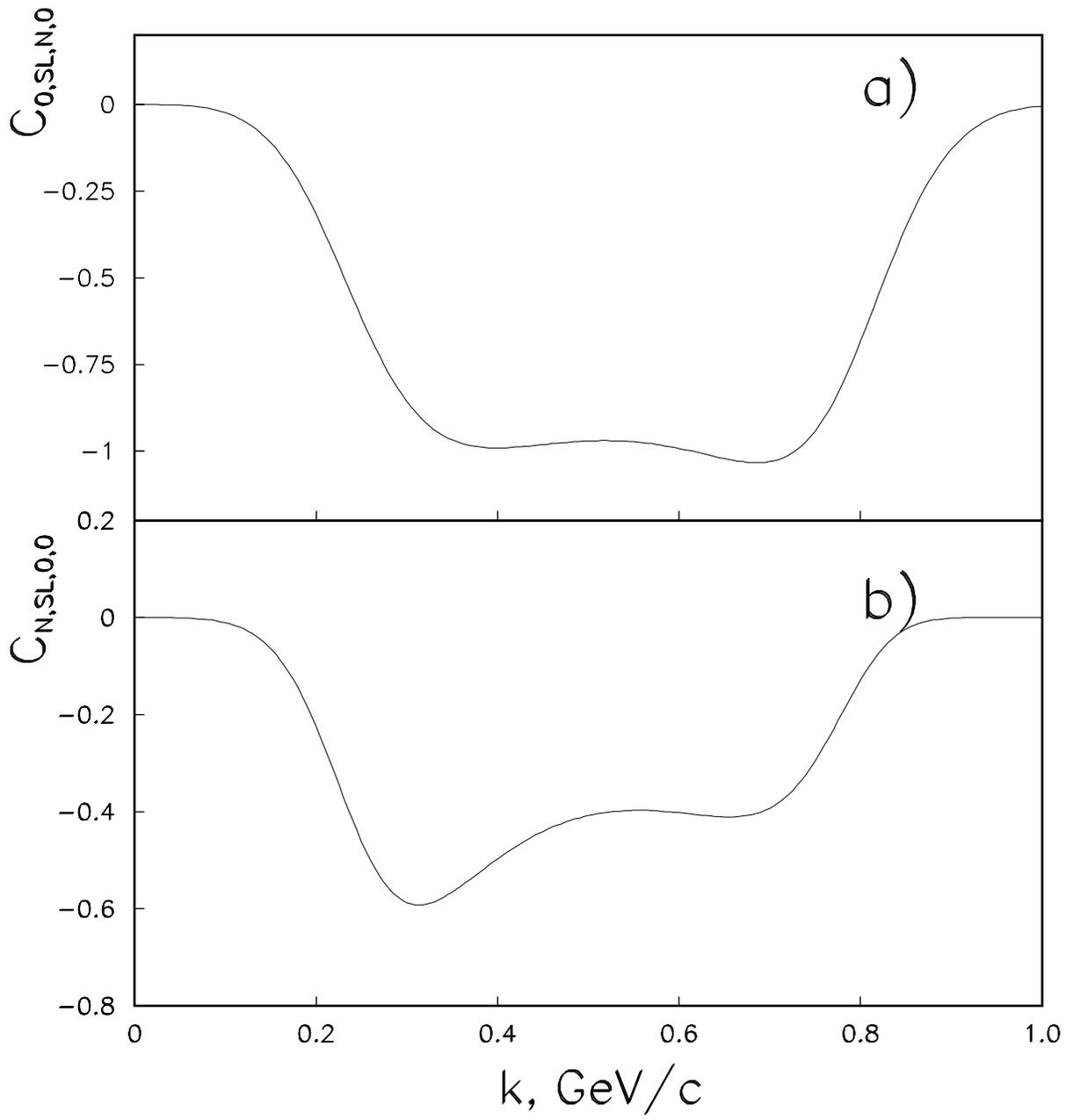
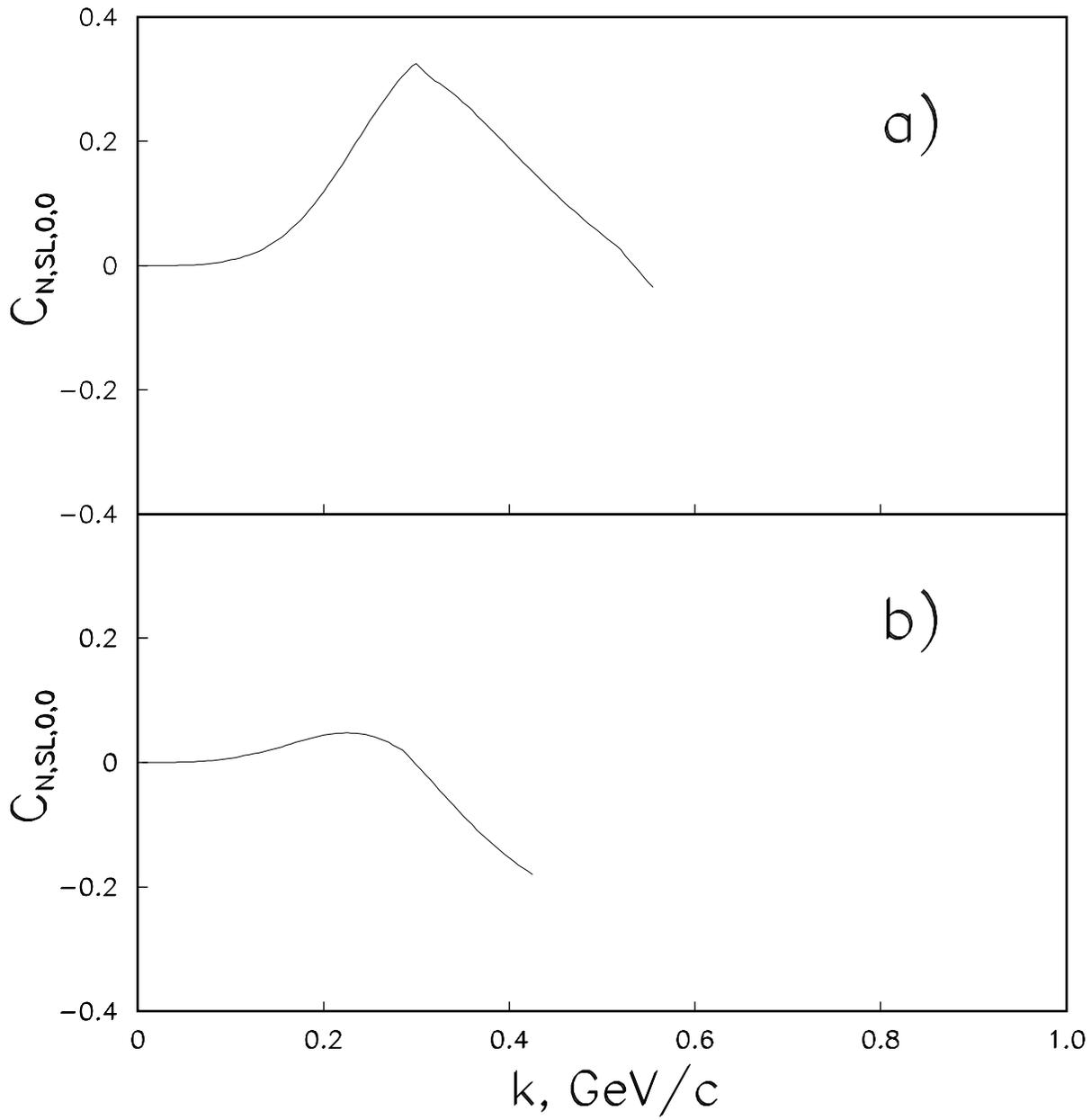
\end{document}